\documentclass[prl,letterpaper,twocolumn,final,superscriptaddress,floatfix]{revtex4}
\usepackage{graphicx,psfrag,amsmath,amssymb,amsfonts,bbm,latexsym,color,dcolumn,bm,mathbbol}
\usepackage[framemethod=tikz]{mdframed}
\usepackage{hyperref}
\usepackage{xfrac} 
\usepackage{ifthen}
\usepackage{xcolor}
\usepackage{textgreek}

\usepackage{amsmath,amssymb}             
\usepackage[french,english]{babel}

\newcommand{\refn}[1]{Eq. (\ref{#1})}
  
\newcommand{\pa}{\partial}

\def\p{\partial} 
\def\nn{\nonumber}
\def\f{\frac}
\def\l{\left(}
\def\r{\right)}
\def\la{\langle}
\def\ra{\rangle}

\usepackage{enumerate}

\newcommand{\cverbose}{1}

\begin{document}

\graphicspath{{Fig/}}

\title{Maximal fluctuations of confined actomyosin gels: dynamics of the cell nucleus}

\author{J.-F. Rupprecht} 
\affiliation{Mechanobiology Institute, National University of Singapore, 5A Engineering Drive 1, 117411 (Singapore).}

\author{A. Singh Vishen}
\affiliation{Simons Centre for the Study of Living Machines, National Centre for Biological Sciences (TIFR), Bangalore 560065 (India).}

\author{G. V. Shivashankar}
\affiliation{Mechanobiology Institute, National University of Singapore, 5A Engineering Drive 1, 117411 (Singapore).}

 \author{M. Rao}
\affiliation{Simons Centre for the Study of Living Machines, National Centre for Biological Sciences (TIFR), Bangalore 560065 (India).}

\author{J. Prost}
\email{jacques.prost@curie.fr}
\affiliation{Mechanobiology Institute, National University of Singapore, 5A Engineering Drive 1, 117411 (Singapore).}
\affiliation{Laboratoire Physico Chimie Curie, Institut Curie, PSL Research University, CNRS UMR168, 75005 Paris, France}

\begin{abstract}
We investigate the effect of stress fluctuations on the stochastic dynamics of an inclusion embedded in a viscous gel. We show that, in non-equilibrium systems, stress fluctuations give rise to an effective attraction towards the boundaries of the confining domain, which is reminiscent of an active Casimir effect. We apply this generic result to  the dynamics of deformations of the cell nucleus and we demonstrate the appearance of a fluctuation maximum at a critical level of activity, in agreement with recent experiments [E. Makhija, D. S. Jokhun, and G. V. Shivashankar, Proc. Natl. Acad. Sci. U.S.A. 113, E32 (2016)]. 
\end{abstract}

\date{\today}

\ifthenelse{ \cverbose > 1}{}{\maketitle}

There has been growing interest in the role of intracellular mechanical fluctuations on cell behavior, with several recent studies showing that stem cells and cancer cells display higher fluctuation levels than normal differentiated cells \cite{Talwar2013,Guo2014,Mandal2016a}. The corresponding physiological role of such fluctuations remains unclear. Contrary to expectations, recent experiments have shown that fluctuations of nuclear components (eg. membrane or DNA loci) are not governed by intra-nuclear activity, but rather by the extra-nuclear actomyosin contractility \cite{Weber2012,Hameed2012,Versaevel2012,Makhija2015a, Ramdas2015,Radhakrishnan2017}. 

In this Letter, we introduce a simple model based on active gel theory \cite{Kruse:2005,Marchetti2013, Prost2015} which illustrates how variation of active stress noise can induce a maximum in the fluctuation spectrum of the nuclear shape. 
We discuss the rather unexpected experimental result of Ref. \cite{Makhija2015a} which, combining drug treatments and geometric constraints, shows that the nuclear area fluctuations are maximal for an intermediate level of contractility. Treatment by Cytochalasin-D -- an actin depolymerazing agent -- reduces the amplitude of nuclear fluctuations in cells placed on small circular patches (ie. with low level of contractile filamentous actin) while the same drug increases nuclear fluctuations in cells on large rectangular patches (ie. with a high level of contractile filamentous actin). 

Though we focus here on the cell nucleus, our findings are also applicable to many other situations in which an active medium drives fluctuations of an inclusion, e.g. tri-cellular junctions in a tissues \cite{Curran2017}  or colloids in active suspensions \cite{Lau2009, Vizsnyiczai2017, Marchetti2013}.

We describe the dynamics of an inclusion subject to stress fluctuations arising from the surrounding actomyosin fluid confined within $x \in \left[-L,L\right]$ (for simplicity, we consider a one dimensional situation).  We successively investigate the cases of a rigid inclusion  (eg. a stiff nucleus) and of an elastic inclusion (eg. soft nucleus) that is embedded in a confined active gel of fluctuating activity; further geometries are considered in Ref. \cite{Singh2017}. We find that, in a non-equilibrium situation, stress fluctuations generate an effective attraction towards the edges of the confining domain. Based on a Maxwell description of the gel, we check that this effect does not appear for thermal fluctuations. The attraction reported here is distinct from the well-known accumulation of dry active particles close to walls \cite{Solon2015,Elgeti2015a}, as we consider a passive inclusion and hydrodynamics interactions; it is rather reminiscent of a Casimir effect, since the fluctuation-induced force on the inclusion originates from its surrounding active medium \cite{Bartolo2003,Aminov2015}. We finally predict the existence of an optimal contractility level that maximizes the amplitude of fluctuations of an elastic inclusion, providing a rationale for the unexpected experimental results of \cite{Makhija2015a}. 

We propose an original model for the active stress fluctuations felt by the inclusion (whether rigid or elastic). We expect cortical actomyosin fluctuations to be the leading contribution to the nuclear dynamics since cortical dissipation is significantly larger than cytoplasmic dissipation at relevant length and time scales of the cell \cite{Turlier2014,Rupprechta}. Integrating over the apical cortex thickness, we find that the tension $\sigma(x, t)$ along the gel coordinate $x$ reads:
\begin{align}  \label{eq:constitutive_no_Maxwell}
(\tau_v \pa_t + 1) \sigma(x,t) = \eta \pa_x v(x,t) +  \zeta \Delta \mu +  \theta_A + \phi_T ,
\end{align}
where $\tau_v$ is the Maxwell relaxation time of the gel; $\eta$ is an effective one dimensional viscosity and $v(x)$ is the gel velocity; $\zeta \Delta\mu$ models the medium contractility, e.g. the tension generated by contractile motors \cite{Kruse:2005,Prost2015}; $\phi_T$ and $\theta_A$  are fluctuating tensions of thermal and active origins, respectively. In the cell context, we expect those active fluctuations to result from fluctuations in the cortical thickness (see SM \cite{Rupprechta}).  We assume that these noise sources are mutually independent and spatially uncorrelated. Notice that we consider a constant viscosity in \refn{eq:constitutive_no_Maxwell}; as a consequence of the fluctuation-dissipation theorem, the thermal noise $\phi_T$ is to be considered as delta-correlated in time: $\langle \phi_T(x,t)  \phi_T(x^{\prime},t^{\prime}) \rangle =  2 \Lambda_T \delta_{x,x^{\prime}} \delta_{t,t^{\prime}}$ \cite{Mori1965,Kubo1966}. In contrast, we assume that the active noise $\theta_A$ displays significant time correlation, which we choose to be exponential: $\langle \theta_A(x,t)  \theta_A(x^{\prime},t^{\prime}) \rangle =  \Lambda_A \delta_{x,x^{\prime}} \exp(-\lvert t - t'\lvert/\tau_A)/\tau_A$. Similar assumptions on the active noise have been introduced to study the diffusion of DNA loci in the presence of intra-nuclear remodeling processes \cite{Ghosh2014,Vandebroek2015,Osmanovic2016a}. Here, activity is related to cortical remodeling processes; we estimate that $\tau_A \approx 60 \, \mathrm{s}$ based on images of the dynamics of apical actin foci \cite{Li2014}. 

\begin{figure}[t!]
	\includegraphics[width=8.00cm]{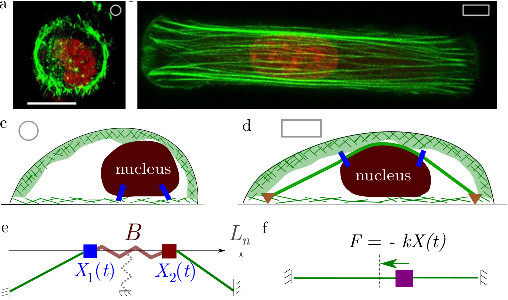}
  \caption{ 
(a, b) Projected intensity of actin (phalloidin labelling) and of the nucleus (DAPI labelling of DNA) of cells placed on (a) circular and (b) rectangular patches of fibronectin (Scale bar: $10 \mu \mathrm{m}$, excerpt from Ref. \cite{Makhija2015a}).
(c, d) Schematic representation of the cortex (green) within cells placed on patches either (c) circular ($500 \, \mu \mathrm{m}^2$) or (d) rectangular ($1800 \, \mu \mathrm{m}^2$). Physical links (blue and red) connect the nucleus (grey) to the actin meshwork (green). 
In (d), the nucleus is compressed by actin fibers (solid green line) anchored to the substrate through focal adhesions (brown triangles). 
(e-f) Models considered here, with (e) a restoring force provided by the geometry of the active segments on an elastic inclusion maintained at a distance from the base and (f) a linear restoring force on a rigid inclusion.}
\label{Fig1}
\end{figure}

We now consider a rigid inclusion at the position $x_t$ (see Fig. \ref{Fig1}f). At the cell scale, low Reynolds number dynamics holds and forces should be balanced at any time step. In the absence of external friction, this implies that tension should be constant in both segments to the left and right of the inclusion, e.g. $\sigma(x,t) = \sigma^{(L)}(t)$ for $x \in \left[-L, x_t\right]$ (conversely $\sigma^{(R)}$ on the right). The existence of a fluctuating gel velocity field leads (by continuity) to the random motion of the inclusion. Averaging \refn{eq:constitutive_no_Maxwell} over the segment $\left[-L, x_t \right]$ leads to
\begin{align} \label{eq:stress}
(\tau_v \pa_t + 1) \sigma^{(L)}  &= \zeta \Delta \mu + \frac{\eta \dot{x}_t }{L+x_t} + \frac{\int^{x_t}_{-L} (\theta_A + \phi_T)}{L+x_t} ,
\end{align}
where we assumed a zero velocity on the edge $-L$. We derive a similar equation on $\sigma^{(R)}$. 

 \begin{figure}[t!]
	\includegraphics[width=8.75cm]{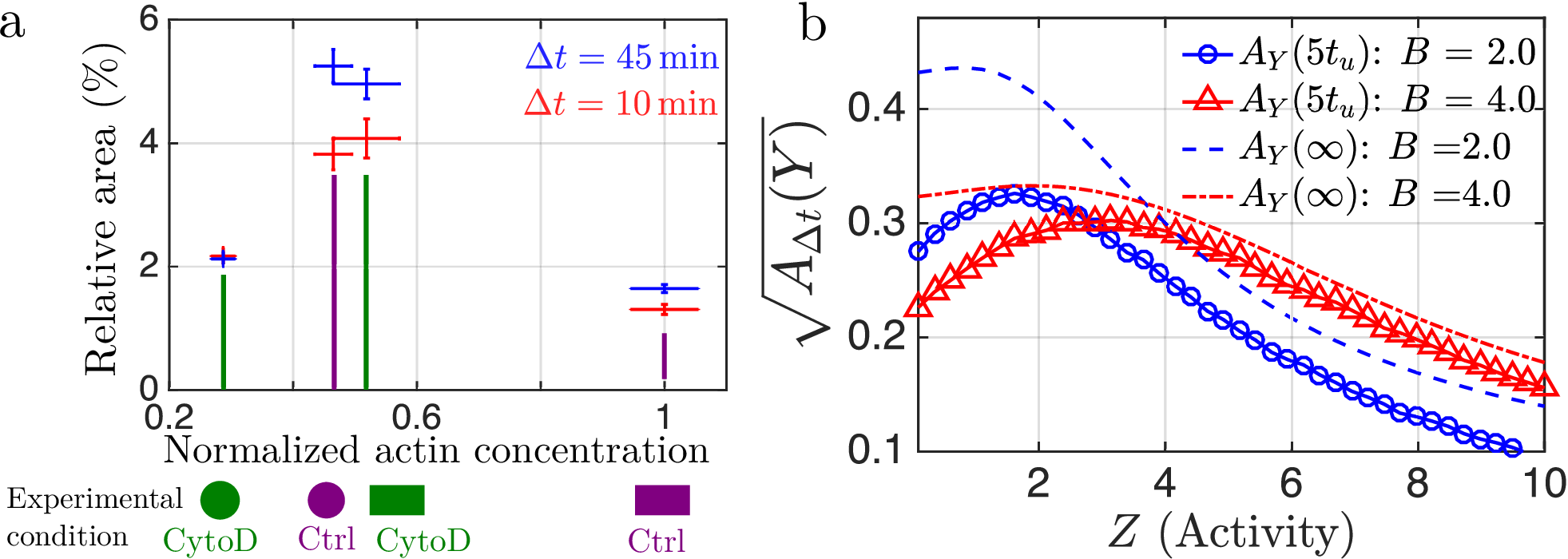}
  \caption{(a) Nuclear fluctuations of the projected area as a function of the cell actin concentration (phalloidin label) normalized to $1$ in the rectangular control case, estimated over time windows. The actin-disruption drug Cytochalasin-D (CytoD) reduces the actin concentration. (b) Finite time variance in the width of an elastic inclusion (defined in \refn{eq:autocorrelation}) as a function of the cortical activity parameter $Z$, for two values of the normalized nuclear elasticity $B = 2$ (blue circles) and $B = 4$ (red triangles) with a model nuclear height $L_n = 0.2 L$. The observation time $\Delta t = 5 t_u = 5 \cdot 10^{3} \, \mathrm{s}$ is comparable to the duration of experiments in \cite{Makhija2015a} (see SM \cite{Rupprechta}). We model the effect of the drug CytoD though a reduction in the normalized activity parameter $Z$.}
\label{Fig2}
\end{figure}

Combining \refn{eq:stress} with the force balance equation on the inclusion $\sigma^{(L)} - \sigma^{(R)} = f(x_t) - m \dot{v}_t$, where $f$ is an externally applied force and $m$ is the mass of the inclusion, leads to the following dynamics on $v_t = \dot{x}_t$:
\begin{align} \label{eq:firsteq_Langevin}
\tau_v \ddot{v}_t +  \dot{v}_t + \frac{\lambda(x_t)}{m} v_t = 
\frac{f  + \tau_v \dot{f}  +  \mu(x_t) \left[\Theta_A + \Phi_T\right]}{m} ,
\end{align}
where $\Theta_A$ is an exponentially correlated noise ($\left\langle \Theta_A(t) \Theta_A(s) \right\rangle = \Lambda_A \exp(-\lvert t-s \lvert/\tau_A)/\tau_A$) and $\Phi_T$ is a Gaussian white noise of variance $2 \Lambda_T$; the friction term reads $\lambda(x)  = (2 \eta L)/(L^2- x^2)$ and
\begin{align} \label{eq:gA}
\mu^2(x) = \frac{2 L}{L^2-x^2} = \frac{\lambda(x)}{\eta}.
\end{align}

Here, we derive the Fokker-Planck equation corresponding to \refn{eq:firsteq_Langevin} through successive adiabatic elimination of the time scales $\tau_m = m/\lambda(x) \ll \tau_v$ and $\tau_v \ll t_u$, where $t_u = \eta^2 L/\max(\Lambda_A,\Lambda_T)$ is a characteristic displacement time of the nucleus. Our method generalizes Ref. \cite{Sancho1982}, which did not include the fluid relaxation time. We first notice that the long-time solution of \refn{eq:firsteq_Langevin} can be expressed in terms of a Green function $\chi^{(t-s)}_{x_t}$ as 
\begin{align}  \label{eq:newton}
m v(t) = \int^{t}_{-\infty} ds \, \chi^{(t-s)}_{x_t} \left\lbrace f   + \mu(x) \left[\Theta_A + \Phi_T\right] \right\rbrace,
\end{align}
where the term $\dot{f} = f^{\prime} v_t$ is included in the friction term. At first order in $\tau_m \ll \tau_v$, we find that:
\begin{align} \label{eq:def_chi}
 \chi^{(t-s)}_{x_t} = \sqrt{\frac{\tau_m}{\tau_v}} \sin \left[\frac{t-s}{\sqrt{\tau_m\tau_v}} \right] \exp\left(-\frac{t-s}{2 \tau_v}\right).
\end{align}
The integral on the continuous functions $f$ and $ \mu \Theta_A$ can be simplified thanks to the relation $ \chi^{(t-s)}_{x_t} \sim \tau_m \delta_{t,s}$ in the limit $\sqrt{\tau_m \tau_v}  \ll t-s$. However, this relation does not hold over the discontinuous white noise $\mu \Phi_T$. Following \cite{Sancho1982}, we consider the Taylor expansion of the noise amplitude $\mu(x)$ to obtain:
\begin{align} 
\mu (x_s) &= \mu(x_t) - \frac{\mu^{\prime} (x_t) \mu(x_t)  }{m} \int^{t}_{s} \!\!\! \int^{t^{\prime}}_{-\infty} \! \! \! 
 \chi^{(t^{\prime}-t^{\prime \prime})}_{x_t} \Phi_T(t^{\prime \prime}), \label{eq:mu_expansion}
\end{align} 
which holds up to $ \tau^{1/2}_m $ terms. From Eqs. (\ref{eq:newton}) and (\ref{eq:mu_expansion}), we obtain the following Langevin equation that is valid at $\tau^{3/2}_m$ order:
\begin{align}
v_t = \frac{\tau_m}{m}  \left\lbrace f +  \mu \left[ \Theta_A  + \Theta_T \right] \right\rbrace  + \Omega_T,   \label{eq:velocitysdeorder3} 
\end{align}
where  $\Theta_T= (1/\tau_m) \int^{t}_{-\infty} ds \,  \chi^{(t-s)}_{x_t}  \Phi_T(s)$ is an exponentially correlated noise  and
\begin{align}
\Omega_T &= \frac{\mu \mu^{\prime}}{m^{2}} 
\int^{t}_{-\infty} \chi^{(t-t^{\prime})}_{x_t} \Phi_T(t^{\prime})
 \int^{t^{\prime}}_{t} \!
\int^{t^{\prime \prime}}_{-\infty} \, \chi^{(t^{\prime \prime} -t^{\prime \prime \prime})}_{x_{t^{\prime \prime}}}  \Phi_T(t^{\prime \prime \prime}). \nonumber  
\end{align}
After some algebra, we find that $\left\langle \Omega_T \right\rangle = - \tau^2_m/(2 m^{2}) \mu^{\prime}  \mu$ and $\left\langle \Omega_T(t) \Omega_T(s) \right\rangle = 0$ at order $\tau^{3/2}_m$. This leads to the following dynamics on $\dot{x}_t  = v(t)$ at order $\tau^{3/2}_m$
\begin{align} \label{eq:FKf}
\dot{x}_t = \frac{f}{\lambda}
- \frac{\Lambda_T \mu \mu^{\prime} }{2 \lambda^{2}} 
+ \frac{\mu \left[\Theta_A 
+ \Theta_T\right]}{\lambda},
\end{align} 
where $\Theta_{A/T}$ are exponentially correlated noise of variance $\Lambda_{A/T}$. In the limit $\tau_v  \ll \tau_A \ll t_u$, both $\Theta_{T}$ and $\Theta_{A}$ are to be interpreted in the Stratonovich convention \cite{Wong1965,Gardiner2009} and the Fokker-Planck equation associated to \refn{eq:FKf} reads: 
\begin{align} \label{eq:fokker_planck_1}
\pa_t  P &= \partial_x  \left\lbrace \left[ -\frac{f }{\lambda }  + \frac{\Lambda_T \mu \mu^{\prime}}{2 \lambda^2 } \right] P + \frac{\Lambda \mu}{2 \lambda } \partial_x  \frac{\mu}{\lambda }  P  \right\rbrace,
 \end{align}
 where $\Lambda = (\Lambda^2_T + \Lambda^2_A)^{\sfrac{1}{2}}$. 

In the absence of an active noise ($\Lambda_A=0$), we find that the steady state solution of \refn{eq:fokker_planck_1} is the Boltzmann distribution $P  \propto \exp(-U(x)/k_b T)$, where $f = - \pa_x U$ and  $k_b T = \Lambda_T/\eta$.
The dynamics  $\dot{x}_t = f/\lambda + (\mu/\lambda) \Theta_T$ is consistent with thermodynamics only when the Gaussian white noise is interpreted with the Hanggi-Klimontovich convention \cite{Lau2007,Sokolov2010}.

Whenever $\Lambda_A>0$, the steady state of  \refn{eq:fokker_planck_1} is not Boltzmann distributed and there is no effective temperature such that $P  \propto \exp(-U(x)/k_b T_{\mathrm{eff}})$. Remarkably, \refn{eq:fokker_planck_1} shows that, in general, the coexistence of thermal and active noise sources cannot be described by any $\alpha$-convention \cite{Lau2007}, e.g. neither by the Stratonovich ($\alpha = 1/2$) nor by the Hanggi-Klimontovich ($\alpha = 1$) conventions.

When the thermal noise is negligible ($\Lambda_T = 0$), \refn{eq:fokker_planck_1} amounts to
\begin{align} \label{eq:Strato}
\partial_t P &=  \partial_x \left\lbrace - \frac{f}{\lambda} P  + \Lambda_A \frac{\mu}{2 \lambda} \partial_x  \frac{\mu}{\lambda}  P  \right\rbrace,
\end{align}
which corresponds to the Stratonovich convention of \refn{eq:FKf}.
This result originates from our assumption that the active noise is correlated over a long time scale compared to the viscous memory time ($\tau_A \gg \tau_v$).



\begin{figure}[t!]
	\includegraphics[width=7.25cm]{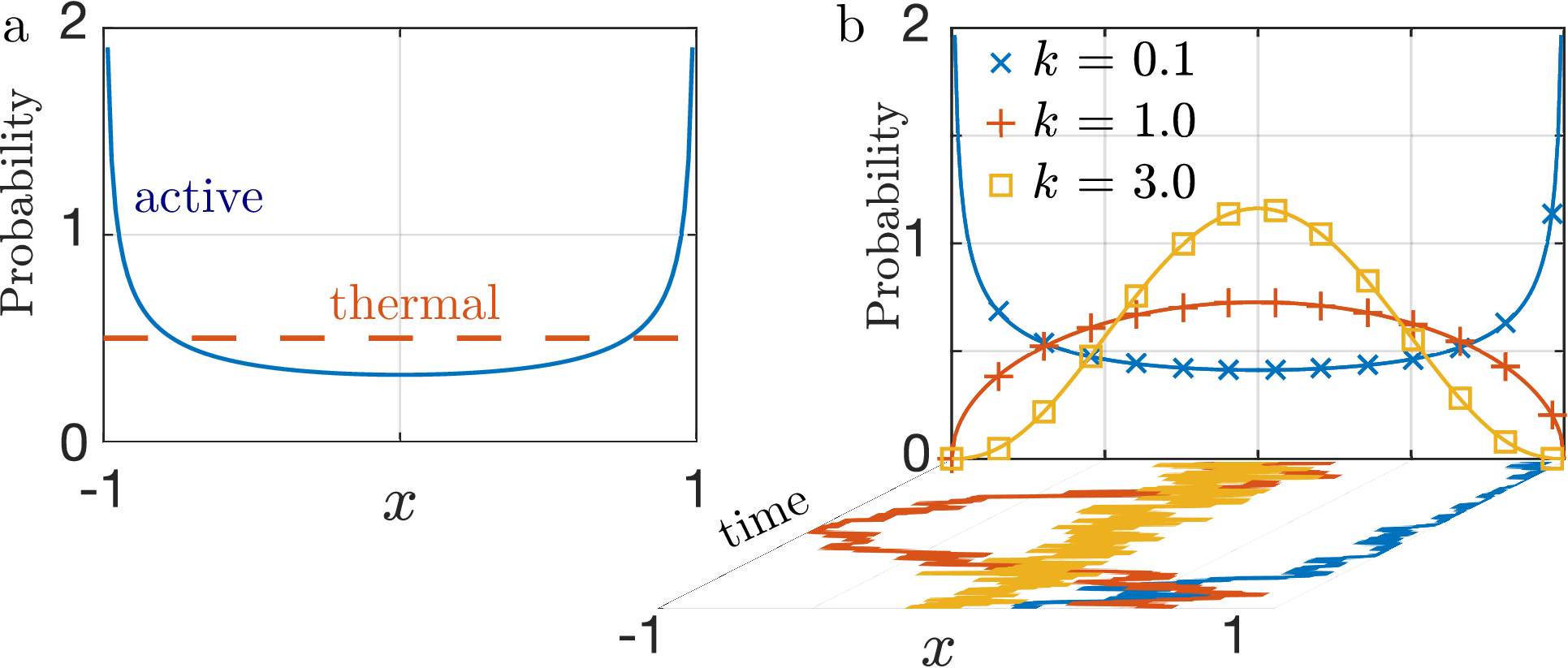}
  \caption{Probability distribution (a) for an active noise (solid blue) and for a thermal noise (dashed red) in the absence of an external force; (b) for an active noise and for various strength of the restoring force $k = 0.1$ (blue), $k =1$ (red) and $k = 3$ (orange). (b, $xy$ plane) Typical trajectories over $\Delta t = 2$: for $k=0.1$ (blue), the inclusion is confined to the edges, while for a large restoring force $k = 3$ (orange), it is confined to the center.  For $k=1$ (red), the particle exhibits the largest displacements over the observation time $\Delta t$.}
\label{Fig2traj}
\end{figure}

We now apply \refn{eq:Strato} in the case of a harmonic restoring force $f(x) = -K x$; additionally, we assume a constant friction $\Gamma(x)= L/(2 \eta)$ for illustration purposes (see \cite{Singh2017} for the complete case). Then, the stationary probability distribution reads
 \begin{align} \label{eq:general_solution_withforce}
P(x) = \frac{\sqrt{\pi } \Gamma \left(k+\frac{1}{2}\right)}{L \Gamma (k+1)}  (1-(x/L)^2)^{-1/2 + k},
\end{align}
where $k = (K L^2 \eta)/(2 \Lambda_A)$ is a dimensionless parameter.
From \refn{eq:general_solution_withforce}, we observe a significant change in the shape of the probability distribution at $k = k_c = 1/2$. For large restoring forces $k > 1/2$, the inclusion is localized around the center. In contrast, for small restoring forces $k < 1/2$ the probability distribution diverges on the edges, indicating the inclusion is attracted by the boundaries (see Fig. \ref{Fig2traj}). This divergence can be traced back to the fixed velocity boundary condition at the edges which leads to a diverging friction term (see \refn{eq:gA}).


\paragraph{Finite-time variance}  Even though the shape of the probability distribution undergoes a dramatic change at $k = k_c$, the variance $\mathrm{Var}(x_t)$ of the inclusion position exhibits no singularity as a function of $k$ but rather decreases monotonically as $1/k$. However, we show that fluctuations measured on sufficiently short finite-time windows $\Delta t$ are maximal at an optimal restoring force parameter $k_{\mathrm{opt}}$. We define the following measure of the amplitude of fluctuations, called finite-time variance:
\begin{align} \label{eq:autocorrelation}
A_{\Delta t}(x) = \langle \overline{x_t^2} - \overline{x_t}^2 \rangle_s, 
\end{align}
where $\overline{x}(t) = \int^{\Delta t}_{0} dt^{\prime}\, x_{t^{\prime}}$ refers to a running average over the time window $\Delta t$ while $\left\langle . \right\rangle_s$ corresponds to a steady-state average of the initial position $x_0$. For the dynamics that corresponds to \refn{eq:general_solution_withforce}, we find that
\begin{align} \label{eq:autocorrelation_multiplicative}
A_{\Delta t}(x)  &= \f{1 - \f{2}{(k + 1/2) \Delta t} + \f{1 - e^{-(k + 1/2) \Delta  t}}{(k + 1/2)^2 \Delta  t^2}}{2(k + 1)},
\end{align}
which holds for arbitrary values of $\Delta t$ \cite{Rupprechta}. From \refn{eq:autocorrelation_multiplicative}, we find that $A_{\Delta t}$ exhibits a maximum at an optimal restoring force parameter $k_{\mathrm{opt}}$ as soon as  $\Delta t < \Delta t_c \approx 6.1 t_u$, and that  $k_{\mathrm{opt}} \approx k_c$ for $\Delta t = t_u$  \cite{Rupprechta}. This optimization of fluctuations is illustrated on individual trajectories in Fig. \ref{Fig2traj}b.  For a long observation time $\Delta t \gg t_u$ there is no optimal restoring force, since the finite-time variance converges to $\mathrm{Var}(x_t)$, which decreases monotonically with $k$.  

\paragraph{Geometry-induced centering} As visible on Fig. \ref{Fig1}b, cells placed on rectangular patches display actin fibers on top of the nucleus. We model the nucleus as an elastic inclusion under a contractile compressive loading. We assume that that the two ends of the inclusion remain at a fixed height $z = L_n$. Mechanical equilibrium requires the identity of the projected tensions:  $\sigma_L (1+X_1)/l_t = \sigma_R (1-X_2)/r_t = B(X_2 - X_1)$, where $X_i$ are the coordinates of the two edges of the inclusion (in cell length $L$ unit) while $l =\sqrt{(1+X_1)^2+L^2_n} $ and $r = \sqrt{(1-X_2)^2+L^2_n}$ are the left and right actin fibers lengths (see Fig. \ref{Fig1}e). Applying the same method as in our first calculation, we find that the dynamics of $\bm{X} = (X_1, X_2)$ reads $\dot{\bm{X}} = \bm{A} +  \bm{D} \cdot \overline{\Theta}_A$, 
\begin{align} \label{eq:elastic_ab}
\bm{A} = \begin{bmatrix}
         \frac{ B(X_2 - X_1)  l^3}{(1+ X_1)^2} -\frac{Z \, l^2}{1 + X_1} \\
        \frac{B(X_1 - X_2) r^3}{(1- X_2)^2} + \frac{Z \, r^2}{1 - X_2}
     \end{bmatrix}, 
\bm{D} = \begin{bmatrix}
        \frac{l^{3/2}}{1+ X_1} & 0\\
        0 & \frac{r^{3/2}}{1- X_2}
     \end{bmatrix},
\end{align}
where (i) time is expressed in units of $t_u$, (ii) $B = E L^2 t_u/\eta$ is a normalized elastic modulus, (iii) $Z = \zeta \Delta \mu t_u/\eta$, 
is a normalized apical cortex activity and (iv) $\overline{\Theta}_A = (\Theta_1,\Theta_2)$ is an uncorrelated Gaussian white noise vector with $\left\langle \overline{\Theta}_i(t) \overline{\Theta}_j(t^\prime) \right\rangle  = \delta_{ij} \delta(t- t^\prime)$ \cite{Rupprechta}. 

\begin{figure}[t!]
	\includegraphics[width=7.0cm]{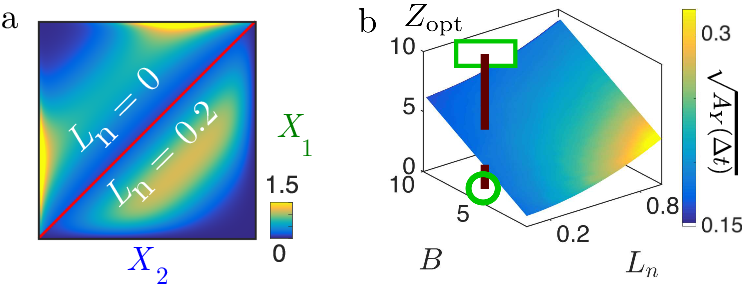}
  \caption{(a) Probability map $P(X_1,X_2)$ for the geometry depicted in Fig. \ref{Fig1}e; active stress is $Z = 3.5$, elastic modulus is $B = 5$ and nuclear height is $L_n = 0$ (resp. $0.2$) in the upper (resp. lower) half. Activity tends to confine the nucleus to the edges of the gel, while the nuclear height focuses the nucleus to the cell center. (b) Surface of the optimal activity level $Z_{\mathrm{opt}}$, colored by the value the maximal $\sqrt{A_{\Delta t}(Y)}$. The red line corresponds to the set of parameters used in Fig. \ref{Fig2}b.
}  
\label{Fig3}
\end{figure}

We interpret \refn{eq:elastic_ab} in the Stratonovich convention under which the Fokker-Planck equation reads: $\pa_t P = \pa_i \left( A_i P + D_{ik} \pa_j D_{jk} P \right)$, with summed indices \cite{Gardiner2009}. The corresponding stationary probability distribution reads:
\begin{align} \label{eq:joint_probability}
P(x_1,x_2) = \frac{N \, (l r)^{3/2} e^{-B (x_2-x_1)^2 - 2 Z (l+r)}}{(1-x_2)(1+x_1)},
\end{align}
where $N$ is a constant \cite{Rupprechta}. The distribution $P(x_1,x_2)$ is peaked at the edges (where fluctuations are minimal) for large value of $Z$ while it is peaked at the center (where fluctuations are maximal) for large values of $B$ (see Fig. \ref{Fig3}a); therefore, we expect fluctuations to be maximal for intermediate values of $B$ and $Z$. For a rigid inclusion ($B = \infty$) and in a flat configuration ($L_n = 0$), \refn{eq:joint_probability} reduces to \refn{eq:general_solution_withforce}.

As a proxy for the nuclear area, we consider the length of the inclusion $Y = X_2 - X_1$.  We evaluate $A_{\Delta t}(Y)$ by Monte-Carlo sampling and we find that, for any given value of $B$ and $L_n$, $A_{\Delta t}(Y)$ is maximal for an optimal activity $Z_\mathrm{opt}$ -- and this even for remarkably large values of the observation time window $\Delta t = 5 t_u$. Simulations further indicate the finite time variance at optimality is an increasing function of both the elastic strength $B$ and of the nucleus height $L_n$; the latter modulates the contribution of the cortical activity to the centering force.

\paragraph{Comparison to experiments}  Our theoretical model explains the counter-intuitive experimental observation that the same biochemical perturbation can either reduce or increase the amplitude of fluctuations, depending on the cell geometry (Fig. \ref{Fig2}a). We expect the cell contractility to be higher in rectangular cells (i.e. large $Z$) than in circular cells (i.e. low $Z$). Indeed, actin filaments appear disorganized in circular confinement while cell-size linear actin structures (called stress fibers) appear in rectangular confinement (see Fig. \ref{Fig1}b). The ordering of myosin motors within stress fibers is known to be similar to that of muscle sarcomeres, which are designed to produce large contractile forces \cite{Burridge2013}. Conversely, the cell cortex produces lower overall contractions since these result from an average over the random organization of acto-myosin structures \cite{Liverpool2008,Lenz2014}. Experiments show that the stress exerted by these fibers lies in $10\, \mathrm{kPa}$ range \cite{Plotnikov2012,Trichet:2012vn}, which leads to a total tension in the $10 \, \mathrm{nN}$ range; in contrast, the apical cortex contractile stress is of the order of $\zeta \Delta \mu_b = 1 \, \mathrm{kPa}$ \cite{Joanny2009}, which leads to a tension in the $1 \, \mathrm{nN}$ range (corresponding to $Z = 1$ \cite{Rupprechta}). Such increase of the cell contractility due to stress fibers  is further supported by the observation that, in rectangular confinement, the nucleus is strongly compressed vertically \cite{Versaevel2012,Makhija2015a}. 

 By varying the contractility $Z$, we reproduce the observation of a maximum in the amplitude of fluctuations in the inclusion size.  The experiments done on rectangular patches with untreated cells correspond to a point on the theoretical curve that is on the right of the maximum, since the contractility is high; reducing contractility by CytoD treatment drives the cell in the direction of the maximum and increases fluctuations (see Fig. \ref{Fig2}a). Conversely, the case of untreated cells on circular patches corresponds to a point close to the maximum; CytoD treatment drives the cell away from the maximum and decreases fluctuations.  We illustrate this on Fig. \ref{Fig3}b  representing the optimal activity maximizing $A_{\Delta t}(Y)$ as a function of the nucleus elastic strength and height. We set $B = 5$, which corresponds to the elastic modulus of the nucleus when probed the $t_u = 10^3 \, \mathrm{s}$ timescale, in the range $E = 10^2 \, \mathrm{Pa}$  \cite{Dahl2005}.

\paragraph*{Conclusion} In this Letter, we show that the incorporation of an out-of-equilibrium fluctuating noise in generalized hydrodynamics equations leads to the attraction of a confined inclusion to the boundaries. In particular, we have shown how active fluctuations lead to steady states that are qualitatively different from any Boltzmann distribution, hence hampering the definition of an effective temperature. This effect explains an experimental observations on the nuclear area fluctuations \cite{Makhija2015a}. 

The fluctuation-induced force described here is generic to any active materials with space-dependent friction and diffusivity. 
Spatially varying friction and diffusion coefficients are commonly encountered in thermal systems that break translational invariance \cite{Reimann:2002ve}, e.g. thermal colloids diffusing near walls  \cite{Lau2007,Faucheux1994}. We expect our result to hold in such two or three dimensional media  since the multiplicative nature of the noise results from an integration over space of field fluctuations. The fluctuation-induced force described here can affect the interpretation of particle tracking experiments, e.g. micro-rheology based on correlations between tracers \cite{Levine2000,Hameed2012} or Bayesian inference of binding potentials \cite{Beheiry2015}. Finally, active fluctuations in tissues may provide unsuspected driving forces for cell neighbour exchange \cite{Curran2017}. These topics will be the subject of our future research work.

\acknowledgments{We thank E. Makhija, D. S. Jokhun and A. V. Radhakrishnan for helpful discussions and comments on the manuscript, as well as F. Julicher, K. Vijaykumar and S. Ramaswamy for fruitful discussions on the thermodynamic limit. J-F. R. and AS thank NCBS Bangalore and MBI Singapore, respectively, for hospitality. J.-F. R., G. V. and J. P. are supported by the National Research Foundation, Prime Minister’s Office, Singapore and the Ministry of Education under the Research Centres of Excellence programm.}

%

%

\clearpage

\onecolumngrid

\appendix

\setcounter{equation}{0}
\setcounter{figure}{0}
\renewcommand{\theequation}{S\arabic{equation}}
\renewcommand{\thefigure}{S\arabic{figure}}

\begin{center}
\Large{Supplemental Material}
\end{center}

\section{Applicability of the model to the dynamics of the nucleus} \label{sec:derivation_constitutive}

\subsubsection{Derivation of the constitutive equation} \label{eq:derivation_constitutive}

\begin{figure}[h!]
  \includegraphics[width=10cm]{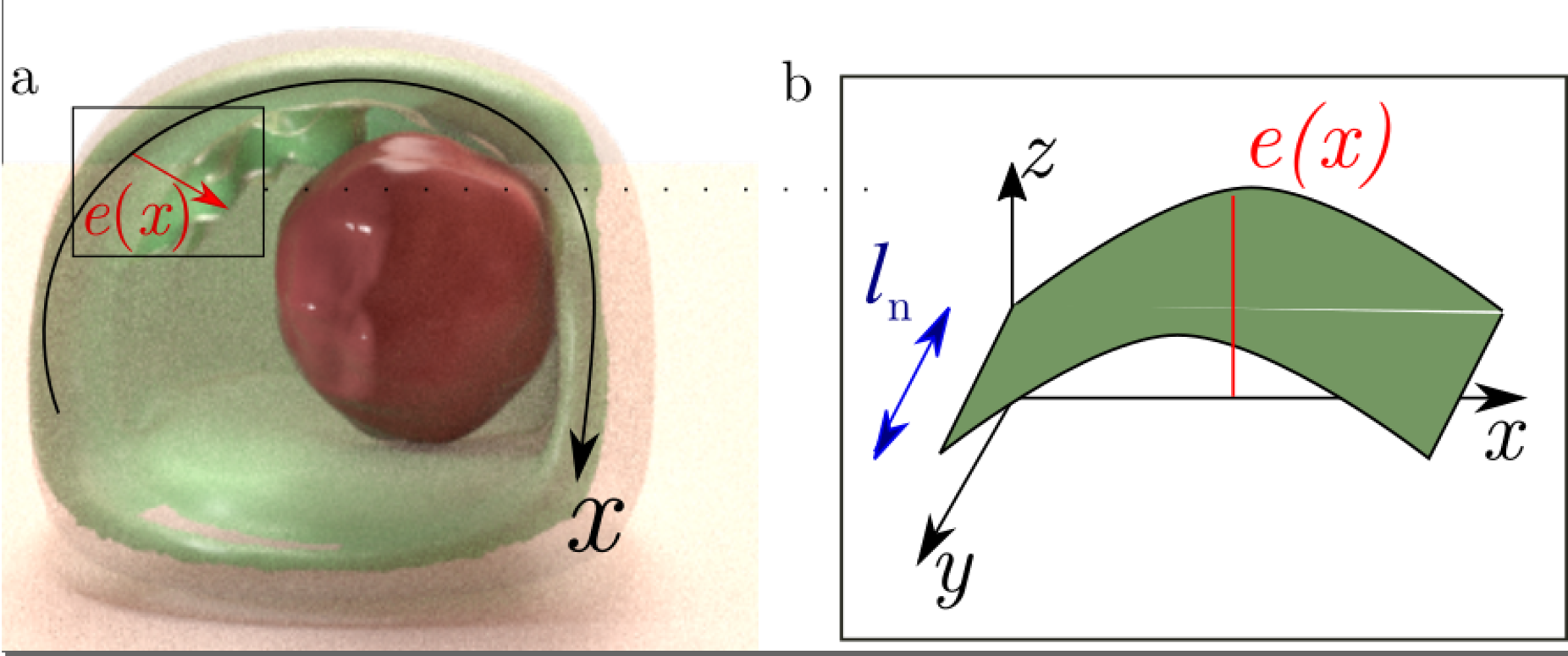}
  \caption{(a) Schematic view of the fluctuating cell cortex (green membrane) and of the nucleus (red sphere). The coordinate $x$ represents a curvilinear axis along the cell cortex. The apical cortex (top of the cell) usually exhibits stronger fluctuations than the basal cortex (bottom of the cell) \cite{Li2014} (b) Model of the local fluctuations of the cortex height $e(x)$, with a transversal width $l_n$.}
\label{Fig_SI1}
\end{figure}

Here, we detail our model assumptions that lead to Eq.\,(1) presented in the main text. We assume that fluctuations in the height of the cortex give rise to an active noise term that is correlated over a long time scale. We model the cortex as a slab of length $-L< x < L$, width $0 < y < l_n$ and height $0 <z < e(x) = e_0 + \delta e$, where $\delta e \ll e_0$ (see Fig. \ref{Fig_SI1}). We then consider an active gel equation for the stress density $\sigma_{b}$ within an elementary volume of the cortex \cite{Prost2015,Kruse:2005}: 
\begin{align} \label{eq:si:constitutivegenerala}
(1 + \tau_v \pa_t) \sigma_{b} = \eta_b \nabla v + \zeta \Delta \mu_b + \varphi_{b,T},
\end{align}
where $\tau_v$ is the gel Maxwell time; $\eta_b$ is the bulk viscosity of the cortex; $\zeta \Delta \mu_b$ is a bulk activity parameter and $\vartheta_{b,T}$ refers to thermal Gaussian white noise of variance $\lambda_T$. 

We consider a weakly perturbed flow conservation equation on a slab of gel of height $e(x)$:
\begin{align} \label{eq:si:conservation_equation}
\frac{\partial e}{\partial t} +  \mathrm{div}\left[e(x) v(x)\right] = \frac{e(x)-e_s}{\tau_A},
\end{align}
where $\tau_A$ is a characteristic relaxation time; 
$e_s = e_0 + \vartheta_A$ is a fluctuating target cortex width, where $\vartheta_A$ is a Gaussian noise that is exponentially correlated in time: $\left\langle \vartheta_A(x,t) \vartheta_A(x^{\prime},t^{\prime}) \right\rangle = \lambda_A \delta(x-x^{\prime})\exp(-\lvert t-t^{\prime} \lvert/\tau_A)/\tau_A$. We focus on the one-dimensional geometry where $\nabla v = \pa_x v(x)$.  Neglecting the time derivative in \refn{eq:si:conservation_equation} (fast cortex relaxation time) and considering the limit of small perturbations $\delta e = e(x)-e_0 \ll e_0$, we obtain the following height fluctuations equation:
\begin{align} \label{eq:si:constitutivegeneralb}
\delta e(x) =  -e_0 \tau_A \pa_x v + \vartheta_A.
\end{align}
The tension exerted on the elementary slab at $x$ of volume $dx \times L_n \times e(x)$ reads $\sigma = \int^{l_n}_{0} \! dy \int^{e(x)}_{0} \! dz \sigma_{b}(x, y, z)$, which, combined with \refn{eq:si:constitutivegenerala} and \refn{eq:si:constitutivegeneralb} leads to 
\begin{align} \label{eq:si:constitutivegeneral}
(1 + \tau_v \pa_t) \sigma
&= \eta \pa_x v(x) + \zeta \Delta \mu + \phi_T + \theta_A,
\end{align}
where $\phi_T \approx \int^{l_n}_{0} \! dy \int^{e_0}_{0} \! dz \varphi_{b,T}$ is approximated as Gaussian white noise of variance $\Lambda_T = e_0 l_n \lambda_T$; similarly, we define $\Lambda_A = e_0 l_n \lambda_A$ such that $\theta_A$ converges to a Gaussian white noise of variance $\Lambda_A$ in the limit $\tau_A \ll t$. Equation \ref{eq:si:constitutivegeneral} corresponds to Eq. (1) in the main text. Notice that we neglect second order terms in $\delta e(x) \ll e_0$ and that we have defined the following one-dimensional effective (i) activity $\zeta \Delta \mu = e_0 l_n  \zeta \Delta \mu_b$, (ii) noise term $\theta_T = e_0 l_n \vartheta_{b,T}$ and (iii) viscosity $\eta = e_0 l_n (\eta_0 - \zeta \Delta \mu_b \tau_A)$, which we expect to be lower than the bare viscosity $e_0 \eta$ as $\zeta \Delta \mu >0$ for actomyosin systems \cite{Hatwalne:2004}).

\subsubsection{Cortical effects versus bulk effects} \label{eq:cortex}

We explain why the contact of the nucleus with the cortex, though relatively thin, should be the leading contribution to the nuclear fluctuations, dominating over contributions from the bulk of the cell. We first show that (1) cortical and cytoplasmic friction forces are small compared to the corresponding viscous contribution, and (2) the viscous contribution of the cytoplasm is negligible compared to the cortex. For this, we consider a viscous two-fluid description of the cytoplasm and cortex. This is valid, since as shown in \cite{Moeendarbary2013}, elastic terms  are important at short times scales, of order $10 \mathrm{s}$, which is significantly smaller than the timescale of interest, $t > 1 \, \mathrm{min}$. Indeed, we expect the Maxwell time of the cortex to be significantly smaller than the cortex turnover time which is also estimated to be $\tau \sim 1 \, \mathrm{min}$ (see \cite{Salbreux2012vn}). 

A general two-fluid (gel+solvent) Stokes description for both the cortex (cor) and cytoplasm (cyt) is  described by,
\begin{align}
 \label{eq:Stokes}
\lambda_{cor} \left(v_{cor}-v\right) & =  \eta_{cor} \nabla \cdot v_{cor} + \nabla \cdot \sigma_{cor}, \\
\lambda_{cyt} \left(v_{cyt}-v\right) & =  \eta_{cyt} \nabla \cdot v_{cyt} + \nabla \cdot \sigma_{cyt},
\end{align}
where $v_{cor}$ and $v_{cyt}$ are the gel velocities of the cortex and cytoplasm respectively and $v$ is the fluid velocity. The $\lambda$'s and $\eta$'s are the 
corresponding permeation coefficients and viscosities of the gel fractions. The rest of the contributions to the local stress density are clubbed into the $\sigma$'s,
written as a sum of an isotropic pressure and active stress, the former getting contributions from both the gel and the fluid.

To show that gel viscous dissipation dominates over gel friction, we need to show that the length scales of interest are smaller than the hydrodynamic screening lengths,
%
\begin{align} \label{eq:fric}
l^{2}_{cor} = \frac{\eta_{cor}}{\lambda_{cor}}\,\,\,\,\, \mbox{and} \,\,\,\,\, l^{2}_{cyt} = \frac{\eta_{cyt}}{\lambda_{cyt}}.
\end{align}
The friction coefficient $\lambda$ can be estimated by equating it to $\eta/a^2$, where $\eta=10^{-3} \, \mathrm{Pa} .\mathrm{s}$ is the solvent viscosity and $a$, the gel mesh size.  For the cortex,
$a_{\mathrm{cor}} \approx 10 \, \mathrm{nm}$, thus with an $\eta_{\mathrm{cor}} = 10^5 \, \mathrm{Pa}.\mathrm{s}$, we find that $l_{cor} = 100 \, {\mu}\mathrm{m}$. Comparing this to  the scale of the nucleus
$\approx 5\,{\mu}\mathrm{m}$, we conclude that the contribution of the friction of the fluid on the cortex to dynamics of the nucleus, is entirely negligible. For the cytoplasm, if we take $a_{\mathrm{cyt}} \approx 10^2-10^3 \, \mathrm{nm}$, then with an
$\eta_{\mathrm{cyt}} \approx 1 \, \mathrm{Pa} \, \mathrm{s}$, we find that $l_{cyt} = 10\, {\mu}\mathrm{m}$.  However, as we show next, cytoplasmic viscous effects have a negligible contribution to the nucleus dynamics at the timescales of interest.
Our numerical estimates above, allows us to drop the friction term in Eq.\,\ref{eq:Stokes}.
We now show that the contribution of the cytoplasm to the dynamics of the nucleus is negligible compared to the cortex. 
Consider the nucleus as an inclusion separating the cell into two compartments - left (L)/ right (R) - with length scales $l_{\mathrm{L}}$ and $l_{\mathrm{R}}$, respectively. 
In the absence of friction, Eq.\,\ref{eq:Stokes} for each compartment reduces to, $\sigma_{cyt}^{\mathrm{L}} \approx \eta_{\mathrm{cyto}}\dot{l}_{\mathrm{L}}/l_{\mathrm{L}} = \eta_{\mathrm{cyto}} \dot{V}_{\mathrm{L}}/V_{\mathrm{L}}$ where $V_{\mathrm{L}}$ refers to the volume on the left hand side of the inclusion (similar relation for $R$). 
Evaluating this at either boundary of the inclusion (assuming continuity of stress density), we obtain,
\begin{align} \label{eq:piston}
\sigma_{cyt}^{L}-\sigma_{cyt}^{R}
=   \eta_{\mathrm{cyt}}
 \left(\frac{\dot{V}_{\mathrm{L}}}{V_{\mathrm{L}}} - \frac{\dot{V}_{\mathrm{R}}}{V_{\mathrm{R}}} \right),
\end{align}
which corresponds to a dissipation stress due to cytoplasmic flow. 
In terms of the position of the inclusion $X_t$, the volumes can be expressed as $V_{\mathrm{L}} \approx l L_n  (L + X_t)$  and $V_{\mathrm{R}} \approx l L_n  (L - X_t)$, where $L_n$ and $l$ are the $z$ (height) and $y$ (transverse) extensions of the nucleus. After integrating the stress Eq. (\ref{eq:piston}) over the transverse area of the nucleus, the corresponding force contribution from the cytoplasm reads 
\begin{align} \nonumber
\sigma_{cyt}^{L}- \sigma_{cyt}^{R} &\approx  \eta_{\mathrm{cyt}} l L_n \left(\frac{1}{L+X_t} + \frac{1}{L-X_t}\right) \dot{X}_t.
\end{align}
To this we add the contribution coming from the cortex of width $e_0$ (see, main text) to obtain the total stress imbalance from L to R. We then obtain a similar stress equation that is identical to Eq.\,(2)  in the main text, but with a new one dimensional viscosity that is $\eta_{\mathrm{new}} = \eta \left[1 +  \eta_{\mathrm{cyt}} l L_n/\eta\right]$. Thus the effect of the pressure difference appears as  a simple increase in the value of the dissipation coefficient. However this enhancement of the effective dissipation is small,  since $\eta_{cor} e_0 l \gg \eta_{cyt} L_n l$, due to the large value of the cortical 
 viscosity: with $\eta_{\mathrm{cor}} \approx 10^{5} \, \mathrm{Pa} . \mathrm{s}$, $e_0 \approx 0.2 \mu\mathrm{m}$, $l = 5 \mu \mathrm{m}$, we get $\eta_{cor} e_0 l \approx 10^{-7}
 \, \mathrm{N} . \mathrm{s}$, whereas with a cytoplasm viscosity $\eta_{\mathrm{cyt}} \approx 1 \, \mathrm{Pa} \, \mathrm{s}$ and $L_n \approx l$, we get $\eta_{\mathrm{cyt}} L_n l \approx 2 . 10^{-11}  \, \mathrm{N} . \mathrm{s}$. This conclusion is consistent with previous analysis, see \cite{Turlier2014}.

\subsubsection{Estimate of the parameters value}

\paragraph*{Correlation time} We identify the active noise correlation time to a typical remodeling time of the cortex.  Confocal imaging shows that the apical cortex exhibits stronger fluctuations: local kinks within the apical cortex appear and vanish within $\tau_A \approx 10^2 \, \mathrm{s}$  \cite{Li2014}. In comparison, the nucleus moves by around $1 \mu m$ within $15$ minutes (see \cite{Radhakrishnan2017}). This separation of time scales supports the hypothesis that the correlation time of the noise can be described through a Stratonovich interpretation of the Langevin equation. The active correlation time is comparable to the actin cortex turnover time \cite{Salbreux2012vn}, which is necessarily significantly larger than the gel Maxwell time $\tau_v$ ($\approx 5 \, \mathrm{s}$ in \cite{Wottawah2005}).  

\paragraph*{Amplitude of the active noise}  We estimate the mean cortical tension to be of the order of $\zeta \Delta \mu_b \delta e_0 l_n$, where $\zeta \Delta \mu_b$ is bulk active stress  $\zeta \Delta \mu_b$ and $\delta e_0 l_n$ is the cortical cross-section. Estimating a  typical fluctuation of the cortex width to be $\delta e_0 = 20 \, \mathrm{nm}$, we expect that the tension fluctuation amplitude should read $\Lambda_A = (\zeta \Delta \mu_b \delta e_0 l_n)^2 \tau_A \lambda$, where we assume that the spatial correlation length is approximatively $\lambda = 100 \, \mathrm{nm}$. This leads to a typical diffusion coefficient of the nucleus $D_{\mathrm{nucleus}} = \Lambda_A L/\eta^2 \approx 10^{-3} \, \mu \mathrm{m}^{2}.\mathrm{s}$ that is consistent with the values measured in \cite{Radhakrishnan2017} over the range $t \approx 10^2 \, \mathrm{s}$. 

\paragraph*{Nuclear elasticity} The order of the elastic modulus of the cell nucleus is typically $E = 1 \, \mathrm{kPa}$ when probed over the range of seconds \cite{Guilak2000,Lammerding2011}. However, when probed at a longer time scale, the cell nucleus appears softer due to the long-time viscous behavior of some of the elastic structures. In particular, \cite{Dahl2005} reports that the nuclear elasticity scales with the timescale of the deformation $t$ as:
\begin{align}
E(t) = E_0 \left(\frac{t}{1 \mathrm{s}}\right)^{-\alpha},
\end{align}
which is indeed a decreasing function of time as $\alpha \approx 0.30 \pm 0.02 >0$. This result was obtained through micropipette aspiration experiments performed on nuclei that were extracted outside of the cell and over a time range spanning from $1 \, \mathrm{s}$ to $10^{3} \, \mathrm{s}$. Assuming that $E_0=10^{3} \, \mathrm{Pa}$, we find that $E(t = t_u  = 15 \, \mathrm{min}) = 10^{2} \, \mathrm{Pa}$.

\begin {table}[h!]
\begin{center}
\begin{tabular}{|l|l|l|}
  \hline
  Quantity & Meaning & Value  \\
  \hline 
  $L$ & Radius of the cell ($x$ direction)  &  $10 \, \mu \mathrm{m}$\\
  $L_n$ & Apical to basal height of the nucleus (circular patch - $z$ direction) &  $3 \, \mu \mathrm{m}$  \\
  $l_n$ & Transversal width of the nucleus (circular patch -- $y$ direction) &  $3 \, \mu \mathrm{m}$  \\
  $e_0$ & Width of the cortex  & $100 \, \mathrm{nm}$  \\
  $\eta_0$ & Bulk viscosity of the cortex & $10^{5} \, \mathrm{Pa}.\mathrm{s}$ \cite{Joanny2009} \\
  $\zeta \Delta \mu_b$ & Cortical active stress & $10^{3} \, \mathrm{Pa}$ \cite{Joanny2009}  \\
  $\tau_A$ & Correlation time, active fluctuations & $10^{2}\, \mathrm{s}$ \cite{Li2014} \\
  $\tau_v$ & Maxwell time & $5 \, \mathrm{s}$ \cite{Wottawah2005}\cite{Saha2016} \\
  $\eta = e_0 L_n (\eta_0 - \tau_A \zeta \Delta \mu_b)$ & One-dimensional viscosity in the presence  of activity  & $5 \cdot  10^{-7} \, \mathrm{N}.\mathrm{s}$  \\
  $\zeta \Delta \mu = \zeta \Delta \mu_b e_0 L_n$ & Active force & $0.5 \, \mathrm{nN}$ \\
  $\delta e_0$ & Typical fluctuating height of the apical cortex & $20 \, \mathrm{nm}$ \\
  $\zeta \Delta \mu_b \delta e_0 L_n$ & Typical fluctuating active tension & $1 \cdot 10^{-10}  \, \mathrm{N}$ \\
  $\lambda$ 						  & Correlation length of the active noise & $100 \, \mathrm{nm}$ \\
  $\Lambda_A = (\zeta \Delta \mu_b \delta e_0 L_n)^2 \tau_v \lambda$ & Variance of the active noise & $10^{-23} \, \mathrm{N}^{2} \, \mathrm{m} \,\mathrm{s}$ \\
  $t_u = \eta^2 L/\Lambda_A$ & Time unit; characteristic displacement time of the nucleus  & $10^{3} \, \mathrm{s} \approx 15 \, \mathrm{min}$ \\
  $D_{\mathrm{nucleus}} = \Lambda_A L/\eta^2 $ & Typical diffusion coefficient of the nucleus & $10^{-3} \mu\mathrm{m} \, \mathrm{s}^{-1}$ \cite{Radhakrishnan2017} \\
 $E_0$ & Nuclear elastic modulus probed at $1$Hz. & $10^{3} \, \mathrm{Pa}$ \cite{Guilak2000,Lammerding2011} \\
  $E(10 \mathrm{min})$ & Nuclear elastic modulus probed over $10 \, \mathrm{min}$ timescale & $10^{2} \, \mathrm{Pa}$ \cite{Dahl2005} \\
  $K$ & Spring constant for the elastic centering force & $10^{-4} \, \mathrm{N}.\mathrm{m}^{-1}$ \\
  $K L_n$ & Elastic centering force & $0.5 \, \mathrm{nN}$ \\
  \hline
  $k = (K L  t_u)/(2 \eta)$ & Elastic centering force (dimensionless) & $2$  \\
  $B = E L L_n t_u/\eta $ & Elastic spring force (dimensionless) & $5$ \\
  $Z = \zeta \Delta\mu t_u/\eta$ & Normalized activity (dimensionless) & $1$ \\
  \hline
\end{tabular}
\caption {List of notations and parameters estimate.} \label{tab:title} 
\end{center}
\end {table}

\section{Rate of convergence of the finite time variance} \label{eq:rateofconvergence}

Here, we derive analytical expressions for the rate of convergence of the finite time variance in two generic cases: the classical Ornstein-Uhlenbeck process with additive noise and the rigid inclusion case considered in  Eq. (3) of the main text, which corresponds to an Ornstein-Uhlenbeck process with a particular choice of a multiplicative noise. 

We evaluate the evolution the first and second moments in the particle position at the time $t$ after initialization at a position $x_0$, denoted $\left\langle x(t) \right\rangle_{x_0}$ and $\left\langle x^2(t) \right\rangle_{x_0}$. Extending the method presented in \cite{Dechant2011}, we show that these two first moments differ from the following running averages $\overline{x} =\left\langle  I_1(t) \right\rangle_{x_0}/t$ and $\overline{x^2} = \left\langle I_2(t) \right\rangle_{x_0}/t$, where:
\begin{align} \label{eq:definition_I1I2}
\left\langle I_1(t) \right\rangle_{x_0} = \int_0^t dt' \, x(t') \quad \mathrm{and} \quad \left\langle I_2(t) \right\rangle_{x_0}  = \int_0^t dt' \, x^2(t').
\end{align}
In terms of \refn{eq:definition_I1I2}, the finite-time variance defined in Eq. (7) in the main text reads:
\begin{align} \label{eq:autocorrelation_intermsofI1I2}
A_{\Delta t}(X) = \langle \overline{X^2} - \overline{X}^2 \rangle_s = \langle I_2/t - I^{2}_1/t^{2} \rangle_s, 
\end{align}
where $\left\langle . \right\rangle_s$ refers to an average over initial positions $x_0$ chosen according to the steady state distribution.

\paragraph*{Moments} We consider a dynamics in the form $\dot{x} = - k x + \sqrt{2 D (1 - x^2)} \, \zeta$, where $\zeta$ is a Gaussian white noise interpreted in Stratonovich convention.  We consider the probability $P = P(x, t \lvert x_0, 0)$ to reach the point $x$ at the time $t$ from the point position $x_0$ at time $t=0$. The forward Fokker-Planck associated to $P$ reads
\begin{align}    \label{eq:fokker_planck_particular}
\p_t P = \p_x \l (k  + D) x + D \p_x (1 - x^2) \r P.
\end{align}
We consider the following averaged over paths that originates from $x_0$. 
\begin{equation} 
\la x^r \ra_{x_0}(t) =  \int^{1}_{-1} \! \mathrm{d}x \, x^r \, P(x,t\lvert x_0,0), \label{eq:averaging_operator_45}
\end{equation}
which we call first ($r=1$) and second moments ($r=2)$.  Averaging \refn{eq:fokker_planck_particular}, we obtain that:
\begin{align}   \label{eq:moments_equation}
\nn \ \p_t \la x^r \ra_{x_0} &=  - \l (k  + D)r + r(r-1) D\r \la x^{r}\ra +  r(r-1) D \la x^{r-2} \ra   -  r D \left[ x^{r-1} (1 - x^2) P \right]^{1}_{-1}.
\end{align}
The last term in the r.h.s of \refn{eq:moments_equation} can be neglected since $P(x)(1 - x^2)$ tends to zero at $x=\pm1$. Therefore, we obtain:
\begin{align} 
\p_t \la x \ra_{x_0}  =  -(k  + D) \la x \ra_{x_0}, \quad \mathrm{and} \quad \p_t \la x^2 \ra_{x_0}  =-2 (k  +  2 D) \la x^2 \ra_{x_0} + 2 D.
\end{align}
With the initial condition $\la x( 0)\ra_{x_0} = x_0$, the solutions of these equations are
\begin{align} 
\la x(t)\ra_{x_0} = x_0  e^{-(k  + D)t} \quad \mathrm{and} \quad
\la x^2(t) \ra_{x_0} = x_0^2 e^{- 2 \l k + 2 D \r t} + \f{D}{k + 2 D} \l 1 - e^{- 2 \l k + 2 D\r t}\r.  \label{eq:secondmoment}
\end{align}
We now define the operator $\left\langle \right\rangle_s$ corresponding to an average over all possible initial position $x_0$ weighted according to the steady state distribution of \ref{eq:fokker_planck_particular}. Importantly, we notice that $\left\langle x^2_0\right\rangle_s = D/(k+2 D)$. A subtle point is that the following measure of the variance: 
\begin{align} 
\la  \la x^2(t) \ra_{x_0} - \la x(t)\ra_{x_0}^2 \ra_s &= \frac{D}{k+2 D } \left( 1  - e^{-2 (k  + D)t} \r,
\end{align}
does not correspond to the finite time variance defined in \refn{eq:autocorrelation_intermsofI1I2}. This is further discussed in the next paragraph. \\

\paragraph*{Finite time variance} Based on the definition \refn{eq:definition_I1I2}, we solve the equation $\pa_t{\langle I_2(t) \rangle}_{x_0} = \left\langle \pa_t I_2(t) \right\rangle_{x_0} = \left\langle x^{2}(t)\right\rangle_{x_0}$ to obtain:
\begin{align}  \label{eq:I2x0_evolution}
\la I_2 \ra_{x_0}(t) &= D \left(\frac{2 D t}{(2 D+k)^2}+\frac{k t}{(2 D+k)^2}+\frac{e^{-2 t (2 D+k)}}{2 (2 D+k)^2}-\frac{1}{2 (2 D+k)^2}\right)+ \frac{D x^2_0}{2 (2 D+k)} \left(1-e^{-2 t (2 D+k)}\right).
\end{align}
Similarly, we find that the following evolution for the mean running average 
\begin{align} 
\overline{x} = \frac{\la I_1 \ra_{x_0}}{t} &= \frac{x_0}{k + 2 D}  \frac{1-e^{-(k + 2 D) t}}{t} .
\end{align}
which, as expected, converges to $x_0$ when $t\ll1$.

Obtaining the evolution of $I_1(t)$ is more complex. Here, adapting the method of \cite{Dechant2011}, we notice that the vector $(I_1,x)$ satisfies the following Langevin equation: $\pa_t (I_1,x) = (x, - k x + \sqrt{2 D (1 - x^2)} \zeta)$, which leads to the corresponding following Fokker-Planck equation
\begin{align}  \label{eq:FP_joint_Ix}
\p_t P &=  - \p_{I_1} (x P)  + \p_x (k  + 1) x P  + D \p_x^2 (1 - x^2) P,
\end{align}
where $P$ is the probability $P = P(x, I_1, t \lvert x_0, 0, 0)$ to reach a point $(x,I_1)$ after a time $t$ after initialization at the point $(x_0,0)$. Similarly to \refn{eq:averaging_operator_45}, we define the average operator 
\begin{align} 
\la x^r I_1^s \ra_{x_0}(t) &=  \int^{1}_{-1} \! \mathrm{d}x \, \int^{\infty}_{-\infty} \! \mathrm{d}I_1 \,  x^r  I^s \, P(x,I_1,t\lvert x_0,0,0).
\end{align}
Integrating the Fokker-Planck \refn{eq:FP_joint_Ix}, we obtain the following closed set of equations for the two moments:
\begin{align} 
\p_t \la I_1 x\ra_{x_0}  =   \la x^2 \ra  -  (k  + D)\la I_1 x \ra,  \quad \mathrm{and} \quad
\p_t \la I^2_1 \ra_{x_0} =  2 \la I_1 x \ra.   \label{eq:I2}
\end{align}
In the rest of the calculation, we set $D = 1/2$ (main text convention). With the initial condition $\la x( 0)\ra_{x_0} = x_0$, the solutions of Eq. \ref{eq:I2} read
\begin{align} 
\la I_1 x \ra_{x_0} &= \frac{e^{-2 (k+1) t} \left(-4 (k+1) e^{\left(k+\frac{3}{2}\right) t}+(2 k+3) e^{2 (k+1) t}+2 k+1\right)}{(k+1) (2 k+1) (2 k+3)} + x^2_0 \frac{2 e^{-2 (k+1) t} \left(e^{\left(k+\frac{3}{2}\right) t}-1\right)}{2 k+3}, \\
 \la I^2_1  \ra_{x_0} &= \frac{e^{-2 (k+1) t} \left((2 k+3) e^{2 (k+1) t} \left(4 k^2 t+6 k (t-1)+2 t-5\right)+16 (k+1)^2 e^{\left(k+\frac{3}{2}\right) t}-(2 k+1)^2\right)}{(k+1)^2 (2 k+1)^2 (2 k+3)}, \\
&+ x^2_0  \frac{2 e^{-2 (k+1) t} \left(-4 (k+1) e^{\left(k+\frac{3}{2}\right) t}+(2 k+3) e^{2 (k+1) t}+2 k+1\right)}{(k+1) (2 k+1) (2 k+3)}, \label{eq:I2_evolution}
\end{align}
Combining \refn{eq:I2x0_evolution} and \refn{eq:I2_evolution}, we derive an expression of the finite time variance from an initial point $x_0$. The average over steady-state distributed initial position
leads to:
\begin{align} \label{eq:autocorrelation_multiplicative_SI}
A_{\Delta t}(X)  &= \f{1}{2(k + 1)} \left[ 1 - \f{2}{(k + 1/2) \Delta t} + \f{1 - e^{-(k + 1/2) \Delta  t}}{(k + 1/2)^2 \Delta  t^2}  \right],
\end{align}
where $t = \Delta t$ is the observation time. Equation (\ref{eq:autocorrelation_multiplicative_SI}) corresponds to Eq. (7) in the main text.  \\

We analyze the existence of an optimal restoring force in two specific regimes
\begin{enumerate}
\item when the observation time is comparable to $\Delta t \approx t_u < \Delta t_c = 6.20 t_u$. In this case, fluctuations measured are maximal at an optimal restoring force parameter $k_{\mathrm{opt}}$ close to the value of $k_c$. Considering the derivative of \refn{eq:autocorrelation_multiplicative_SI} with respect to $k$, we find that the optimal restoring force is a solution of the equation:
\begin{align} \label{eq:ratio}
e^{-(k+1/2) t} = \frac{(2 k+1)^3 t^2-4 (4 k+3) (2 k+1) t+8 (6 k+5)}{8 (k ((2 k+3) t+6)+t+5)},
\end{align}
By considering the limit $k \rightarrow 0$ of \refn{eq:ratio}, we obtain the following equation on the critical observation time $\Delta t_c$ beyond which there is no optimum
\begin{align} \label{eq:t}
e^{-\Delta t_c}=\frac{\Delta t_c^2-12 \Delta t_c+40}{8 (\Delta t_c+5)}.
\end{align}
This condition is not exactly solvable, but we find an approximate solution is $\Delta t_c \approx 6.20$ though a series expansion around $\Delta t_c = 6$. 

\item in the limit $\Delta t \ll t_u$, the existence of an optimum originates specifically from the multiplicative nature of the noise. We find that $k_{\mathrm{opt}}$ continues to decay as $\Delta t$ is increased. From the Eq. (\ref{eq:ratio}), we obtain the following scaling relation for the optimal restoring spring constant  
\begin{align} \label{eq:asympt}
k_{\mathrm{opt}} \sim \frac{2 \sqrt{2}}{\sqrt{\Delta t}},
\end{align}
which holds in the limit of short observation times $\Delta t \ll 1$. In that regime  $\Delta t \ll t_u$, the origin of the optimality lies in the fact that the centering force tends to localize the particle towards a region where the amplitude of fluctuations is maximal. Over very short time scales $\Delta t \ll 1$, we expect the fluctuations to be governed by the local diffusion coefficient $D(x_0) = \sqrt{1-x_0^{2}}$ for a starting position $x_0$; as the centering force is increased (i.e. increasing $k$), the initial position $x_0$ is (as averaged over the steady state distribution) decreased and $D(x_0) $ increased. Therefore, we intuitively expect that an optimum restoring force always exists even for a large restoring force provided that the observation time windows is sufficiently small.
\end{enumerate}

\paragraph*{Ornstein-Uhlenbeck process} To conclude, we check that there is no optimum of the finite time variance in the case of an Ornstein-Uhlenbeck process with additive noise:  $\dot{X}_t = - k X_t + \sqrt{2}\zeta$ with $\zeta$ a standard white Gaussian noise. Following the method defined in the previous paragraph, we find that the finite state variance reads:
\begin{align}
\la A(t) \ra_{s}  &=  \frac{1}{k} - \frac{2}{k^2 \Delta t} + 2 \frac{1 - e^{-k \Delta t}}{k^3 \Delta t^2} .
\end{align}
In contrast to the rigid inclusion case, in the Ornstein-Uhlenbeck case there cannot be any optimum of the finite time variance as a function of $k$.

\end{document}